\documentclass{sig-alternate}

\usepackage{comment}
\usepackage{color}
\usepackage{listings}
\usepackage{textcomp}
\usepackage{graphicx}
\usepackage[ruled,vlined,algo2e,norelsize,linesnumbered]{algorithm2e} 
\usepackage{color}
\usepackage{multirow}
\usepackage{amssymb}
\usepackage{floatrow}
\newfloatcommand{capbtabbox}{table}[][\FBwidth]
\usepackage[font=small,skip=0pt]{caption}

\begin{document}
\title{Scalable Mining of Daily Behavioral Patterns in Context Sensing Life-Log Data} 

\numberofauthors{3} 
\author{
\alignauthor
Reza Rawassizadeh\\
       \affaddr{Department of Computer Science and Engineering,}\\
       \affaddr{University of California Riverside}\\
       \email{rrawassizadeh@acm.org}
\alignauthor
Elaheh Momeni\\
       \affaddr{Faculty of Computer Science}\\
       \affaddr{University of Vienna}\\
       \email{elaheh.momeni.roochi \\
       @univie.ac.at}
\alignauthor Prajna Shetty\\
	   \affaddr{Department of Computer Science and Engineering,}\\
       \affaddr{University of California Riverside}\\
       \email{prajna.shetty@email.ucr.edu}
}
\maketitle
\begin{abstract}
Despite the advent of wearable devices and the proliferation of smartphones, there still is no ideal platform that can continuously sense and precisely collect all available contextual information. Ideally, mobile sensing data collection approaches should deal with uncertainty and data loss originating from software and hardware restrictions. We have conducted life logging data collection experiments from 35 users and created a rich dataset (9.26 million records) to represent the real-world deployment issues of mobile sensing systems. We create a novel set of algorithms to identify human behavioral motifs while considering the uncertainty of collected data objects. Our work benefits from \emph{combinations of sensors} available on a device and identifies behavioral patterns with a \emph{temporal granularity} similar to human time perception. Employing a combination of sensors rather than focusing on only one sensor can handle uncertainty by neglecting sensor data that is not available and focusing instead on available data. Moreover, by experimenting on two real, large datasets, we demonstrate that using a sliding window significantly improves the scalability of our algorithms, which can be used by applications for small devices, such as smartphones and wearables.
\end{abstract}
\vspace{-0.2cm}
\keywords{Motif, Temporal Granularity, Mobile, Life Logging}
\vspace{-0.2cm}
\section{Introduction}
\setlength{\textfloatsep}{8pt plus 1.0pt minus 1.0pt}
\setlength\floatsep{8pt plus 1.0pt minus 1.0pt}
The proliferation of smartphones and, more recently, wearable devices such as fitness trackers and smart watches equip-ped with sensors, has led to a significant expansion of possibilities to study human behavior. Computing and networking capabilities of these sensor-embedded devices makes them appropriate tools for observing and collecting useful contextual information (mobile sensing). For instance, mobile health, which benefits from mobile sensing, offers the possibility of a shift from treatment to prevention in medical care systems. Researchers show that 69\% of U.S. adults monitor and track their health status and 21\% of them use technology for this purpose \cite{tracking4health}. Unlike wearable devices, which are still quite new in the market, the smartphone platform has benefited from a significant amount of scientific work ranging from public transport navigation \cite{micronav} to well-being \cite{bewell}. Both wearable devices and smartphones are very capable of sensing and collecting the basic patterns of human behavior through contextual information.\\
While simple human behaviors are predictable, at least in aggregate \cite{bursts}, traditional approaches for detecting human behavioral patterns (which are not digital)  are often difficult. However, the advent of mobile devices enables researchers to identify human behavior to an extent that was not previously possible. \\
On one hand, this information collection paradigm should be moved from simple data collection tools to intelligent systems with cognition capabilities \cite{smart2cogphone}. On the other hand, there is still a lack of wide acceptance of mobile sensing applications in real-world settings. \\
There are several reasons for this mismatch of capability and acceptance. First is the resource limitation and lack of accuracy in the collected contextual data, especially with regard to the battery life \cite{funchmobcom}. The size of sensors that are dealing with radio frequency, i.e. bluetooth, WiFi and GPS, affects the quality of their data \cite{cacm} (smaller devices have less accurate data). The next reason, which has been noted but has not been widely explored, is the proximity of the smartphone to users \cite{gettingcloser}. Smartwatches and wearables are body-mounted and thus the proximity problem has been resolved in those devices, but they still suffer from a lack of accuracy \cite{validity}. The third reason is operating system restrictions of mobile devices, which removes background services when the CPU is under a heavy load in order to preserve the battery life. As a result, there is no ideal data collection approach that can sense and record individuals' information 24/7 with no data loss. The \emph{uncertainty} of these data objects is a major challenge that limits the applications that can benefit from them.\\
Existing research \cite{realitymining, lausane} on mobile sensing data has offered tentatively promising results, but does not address the uncertainty that exists in a real-world deployment. These studies employ specific hardware which is known for data quality among users; for example, Reality Mining \cite{realitymining} uses Nokia N6600 and the Lausane data campaign \cite{lausane} uses Nokia N95. Since in the real world there are different phone brands and each device has its own restrictions and specifications in terms of software and hardware, we believe these experiments do not consider all aspects of a real-world deployment. \\
To resolve the data collection uncertainty in mobile sensing data analysis, here we introduce novel algorithms that benefits from the variety of sensors on the device. By leveraging previously-collected data our algorithms can predict human behavior with a \emph{temporal granularity} similar to the human perception of time. First, the location estimation algorithm estimates users' location state. Then, the motif detection algorithms extract users' activities, which have been collected from different device sensors, and create a \emph{user profile} from \emph{behavioral motifs}. A behavior motif is a combination of time stamped sensor/data with a confidence level. For instance, \texttt{\{confidence:55\%; 15:00-16:00; location-state:stationary; call:\#{}951603XXXX\}}, shows a motif with 55\% confidence, it describes that between 15:00-16:00 the user is in stationary state and makes a phone call to the given number. \\ 
In a more technical sense, this research has the following novel characteristics:
\vspace{-0.1cm}
\begin{itemize}\itemsep1pt \parskip0pt \parsep0pt
\item{Realistic Data: We could argue that the dataset we have created is the most realistic life logging dataset created to date in comparison with other mobile sensing datasets, such as \cite{realitymining} and \cite{lausane}. Although these studies provide promising results, their data collection is hardware-specific. We claim our approach is very similar to a real-world deployment for the following reasons: (i) Unlike existing research, our experiment did not hand over specific hardware to participants. We relied on users' Android smartphones, which are different brands with different hardware capabilities and different sensors, and this is a significant challenge for data collection. (ii) We asked volunteers with no reward to participate in our experiment. This presents a drawback in that about 2/3 of participants removed themselves from the experiment, but we managed to finish the experiment with 35 participants.}
\item{Temporal Granularity: Human understanding of time is not precise, unlike digital systems. Our daily behaviors occur in time intervals. For instance, a person does not arrive at work every day at exactly the same time, or eat lunch at exactly the same time every day. There is always a time interval for routine behaviors, even if only a small interval, e.g., five minutes for a precise time scheduled such as a meeting. Therefore, there is a need for flexibility in temporal analysis. We implement this important requirement by introducing a simple, but novel human-centric \emph{temporal granularity} method. Our data analysis and algorithms use this temporal granularity instead of the original timestamp.} 
\item{Heterogeneous Data: A salient advantage of our algorithm is its \emph{semantic independence}, which does not consider the type of the underlying sensor data. This makes the algorithm capable of running in any settings that deal with uncertainty and have multiple source of information. It can use any information source (sensors) that has data with a timestamp, whether a continuous timestamp or discrete timestamp. This demonstrates the reliability of the algorithm and makes it applicable to different problem domains.}
\end{itemize}
\vspace{-0.1cm}
The contributions of our work are as follows: (i) identifying users' location states (moving, stationary, unknown) based on location sensors, (ii) converting digital timestamps to a temporal representation similar to human temporal cognition, and (iii) creating a scalable approach that quantifies human behavior by detecting daily-life behavioral motifs from raw sensor data.  \\
To demonstrate the efficiency and utility of this work, we perform the following evaluations. First, we demonstrate that using a sliding window for identification of motifs significantly improves the execution time performance on both experimental datasets, and it significantly outperforms the baseline. This promising result demonstrates that our algorithms are lightweight enough to be integrated into mobile or wearable devices. 
Second, the proposed algorithm identifies users' motifs for one segment of a day (from 00:00 to 08:00) with high reliability (greater than 0.80 accuracy). The identification of users' motifs for other segments of the day has proven to be more difficult, but the algorithms still perform with substantial accuracy (greater than 0.70 accuracy). In addition, we find that performance varies according to the different values of temporal granularity. We look at six different granularities - 5', 15', 30', 60', 90', and 120' - and we find a temporal granularity of one hour outperforms other granularities. Finally, we clustered users based on their identified motifs (using our motif identification algorithms) and the distribution of motifs among segments of the day. We observe three categories of users, whose activities vary in different segments of a day. This observation enables the system to decide on the best runtime for algorithms, which leads to reducing its resource utilization. \\
The remainder of this paper is organized as follows. First we start by describing the dataset and its characteristics. Then we formalize the problem. Next, we describe the design and implementation of our algorithm; this is followed by the experimental evaluation. Afterward, we explain related work and conclude this paper.
\vspace{-0.2cm} 
\section{Dataset}
\begin{figure*}[htb]
\begin{center}
\hbox{\hspace{-2ex}\includegraphics[scale=0.42]{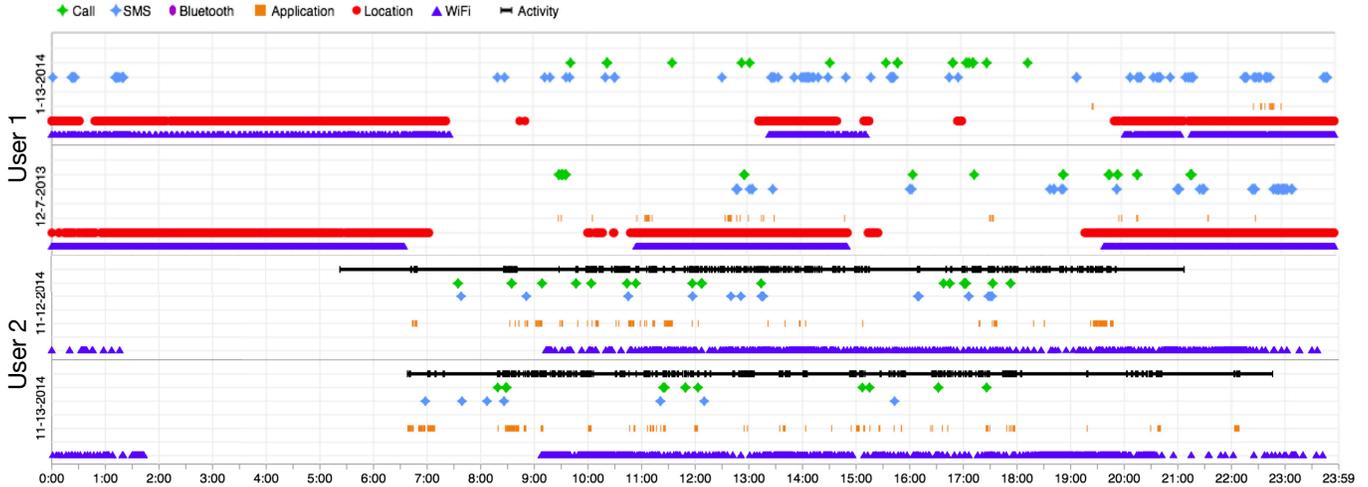}}
\caption{ \small Two day visualization of two different user data. The top two belongs to User 1 and bottom ones belongs to User 2. (best viewed in color) }\label{fig:4dayviz}
\end{center}
\vspace{-0.6cm}
\end{figure*}

As previously noted, the development and testing of our work benefited from access to two real world datasets, Ubiq-Log \cite{ubiqlog} and Device Analyzer \cite{deviceanalyzer}. Unlike previous considered smartphone datasets, i.e. Reality Mining \cite{realitymining} (uses Nokia N6600 ) and Nokia's Lausane data campaign \cite{lausane} (uses Nokia N95), these datasets were collected in real world settings.\\
\textbf{UbiqLog:} We created the UbiqLog dataset. It relied on participants' smartphones and collected a large life log dataset from each participant. We used UbiqLog \cite{ubiqlog} application. Despite the difficulty in doing so, we asked only students who were willing to collect data about their personal lives to participate in our data collection experiment (no other reward). To preserve participants' privacy, UbiqLog is designed in such a way that participants can disable or enable sensors at any time. There were 35 participants, whom 23 are female, with ages ranging from 19 to 32 (Mean= 22.2, SD= 5.6). They collected their data for about two months. We asked participants to enable the following sensors: WiFi, Bluetooth, Location, Application Usage, Call, SMS and Activity (which has been extracted from Google Play Services: on foot, on bicycle, in a vehicle, tilting and still). 
\small
\begin{table}[h]
\begin{center}
\begin{tabular}{ l  p{1.5cm}  p{1.5cm}  }
\hline
Sensor Name   & Num. of Instances & Discarded Instances \\ \hline 
WiFi   &  7640189   & 84 \\ 
Location   &   612901   & 14  \\ 
SMS   & 28486 &  180 \\ 
Call   & 97654 & 7  \\ 
Application Usage   & 753702  & 16  \\ 
Bluetooth Proximity &  117236   &  22 \\ 
Activity State &  15641 &  6 \\ \hline 
All Data Instances  & 9265809  & 329  \\  \hline                  
\end{tabular}
\vspace{-0.2cm}
\caption{ \small UbiqLog dataset records for each sensor.}\label{tab:dataset}
\end{center}
\end{table}
\normalsize
Besides, contact numbers in Call and SMS were stored with pseudonymization and SMS content was completely anonymized. Due to technical difficulties and privacy issues we ran the experiment twice; this paper uses the dataset for the second experiment. \\
Table \ref{tab:dataset} shows a general overview about the data that was collected from participants. Except WiFi and Bluetooth, which were sampled every six minutes, all other sensor data was collected as it became available. There were some corrupt characters in the data, which we ignored at the end due to their insignificance (329 records). Figure \ref{fig:4dayviz} presents four days of data for two users with a visualization we created to gain a high level view of the data.\\
It is notable that all records in the UbiqLog dataset are fine grained information units and semantically rich. In other words, there is no raw sensor data, such as accelerometer data. In summary, there are 9.26 million semantically rich and human readable records.\\
\textbf{Device Analyzer:} The Device Analyzer \cite{deviceanalyzer} has the largest dataset avialable about hardware statuses and device configurations of Android smartphones, collected by the Device Analyzer application. It collects detailed data for about 23,000 users and includes more than 10 billion records of raw sensor data. This is also a promising real-world dataset, but Device Analyzer's focus is on hardware-specific information collection and not user-centric data, unlike our dataset. Therefore, we can not conduct much user behavior analysis, such as motif detection, using this dataset. Nevertheless, we use this dataset to demonstrate the \emph{versatility} of our approach on other context sensing data. For our experiment, we remove hardware-related information, such as network usage, battery state, and system processes. We choose 35 random users to measure execution time performance of our motif detection algorithms. \\
\vspace{-0.5cm}
\section{Definitions \& Problem \\ Statement}
We live in a spatio-temporal world and all of our behaviors occur in a specific location and time \cite{towsharing}. Therefore, a digital system for quantifying human behavior should sense both time and location. Since location sensors such as GPS are not reliable (especially indoor) and it is not possible to collect their data 24/7, we can only use time to link different information together. Human behavior is composed of many daily activities that are distinctive and recurring. These types of activities have been called motifs (or life routines \cite{realitymining}) and our goal is to create a user profile that summarizes the behavioral motifs of a person. \\
We define \emph{entity} as a unit of human activity. It is a tuple of three $e = <T,S,D>$. Each entity contains a timestamp, $T$, sensor name, $S$, and sensor data, $D$. The first task is to find entities that are occurring in the same time interval on different days, based on a given \emph{activity threshold} $\theta$. Therefore, we define the concept of \emph{Group}, $g$ as a set of entities that repeat during a specific number of days, in a specific time interval, i.e., $g=\{e_1,e_2,..., e_m\}$ , $e \in  g$. We can simply compare entities together without creating groups, but to avoid computational complexity we introduce the concept of groups. If we compare entities for all days together, this creates a huge burden on performance $O({2}^{n})$, so to avoid this, we use the sliding window approach. The sliding window reduces the number of comparisons to the size of the window, $m$, and results in windows that can be compared to each other. Therefore, the complexity is $O((n/m)^2 + n/m.(m^2))$, and as $m$ is small, the resulting complexity is $O((n/m)^{2})$. Since the resulting sets are only similar groups and not all entities for a day, the comparison will be reduced significantly and the complexity is $O(n)$. $\theta$ is the minimum threshold for counting similar entities in a specific time interval between days and builds a group. For instance, if $\theta$ is set to three, at least three entities should be repeated in a fixed time interval among a specific number of days. Assuming $T$ (time) is constant among different days, the following equation \ref{eq:group} defines the notation of a group:
\begin{equation}\label{eq:group}
 \ g= \{ \forall e : (e_{i}(T) = e_{i+1}(T) )  \wedge  (\sum _{ i=0 }^{ n }{ e } \geq \theta )\}
\end{equation}
$n$ is the number of entities and it is always smaller or equal to $\theta$ .\\
$B$ denotes \emph{behavior} and is characterized by a set of repeated similar groups $g$ with the same entities among a specific number of days. Each $B$ has at least one group. Therefore, $B=\{g_1,g_2,..., g_n\} $ and $g \subset B$.\\
After behavior $B$ has been created for given dates, the window moves to another set of days and creates another $B$. The second task is to find similar groups that are repeated on all days, with a minimum threshold. In other words, the second step is to find behavioral groups (motifs) that repeat themselves between days. To calculate the similarities between two days we define a threshold, $\lambda$, and name it \emph{motif confidence threshold}. Equation \ref{eqGroup} presents a user profile, which is based on intersections between $k$ number of behavior objects. 
\begin{equation}\label{eqGroup}
	profile = \bigcap_{i=0}^{k} B_i  {\ }, {\ } if (B_i\cap{B_{i+1}}) \geq \lambda 
\end{equation}
At the end we have a single (or multiple if we do the same for weekends or other settings) temporal profile for each user. The profile is composed of users' behavioral groups (motifs) and each identified behavior has a confidence. The confidence presents the probability of the target behavior occurring.
\vspace{-0.1cm}
\section{Proposed Algorithms}
In order to implement algorithms for the problem described above, first the data format should be converted from heterogeneous data to machine processable data. Then we need to convert the timestamp to a time similar to the human perception of time. Furthermore, we propose an algorithm that identifies the location state to enrich the semantic of data with the notion of location. Afterward we describe the behavior similarity and motif detection algorithms. 
\subsection{Data and Location Transformation}
As has been stated before, the data was collected from heterogeneous sources and each sensor provide a semantic rich data element. The following shows a snippet of raw data:
\vspace{-0.1cm}
\small
\begin{verbatim}
{"Application": { 
 	"ProcessName":"com.example.test", 
 	"time":"Oct 15, 2013 6:21:40 AM"}} 
{"SMS": {"Address":"9999999", "Type":"send", 
     "Time":"Dec 24, 2013 11:23:01 PM", 
     "Body":  "anonymized" }} 
{"WiFi": {"BSSID":"f8:d1:38:f4:6b:78", 
     "SSID":"Home", "status":"connected" 
     "time":"Jan 1, 2014 2:09:42 PM" 
     "capabilities":"[WPA2-PSK-CCMP][WPS]"}}  
\end{verbatim}
\vspace{-0.1cm}
\normalsize
These examples show different elements for each record. Therefore, for each sensor we need a unique identifier, we choose ``BSSID'' for WiFi and Bluetooth, the pseudonymized phone number for SMS and Call, ``process name'' for Application and tilting, in-vehicle, on-bicycle, walking, still, unknown for activity sensors (UbiqLog use Google play services for activity recognition and there is no raw accelerometer data inside).\\
As Figure \ref{fig:4dayviz} shows, location data (\textcolor{red}{$\bullet$}) and WiFi data (\textcolor{blue}{$\blacktriangle$}) are not available always. Because of this uncertainty, here location refers to the movement state, which could be \emph{moving}, \emph{stationary} or \emph{unknown}. Our notion of location is more limited than other research efforts which consider the geographical locations or trajectories of users. However, our definition has the twin advantages of simplicity and greater availability. Red dots in Figure \ref{fig:4dayviz}  are not just GPS data. They could be a combination of Cell-ID, GPS, and any third party service that provides geographical coordinates. There are a few works \cite{gpsgsm, enefpossmart} that focus on extracting location from \emph{combination} of different data sources. However, there are several other works that focus on extracting location from a \emph{single} source of information \cite{discgps, minindqu, minprofreq, mineusermob, semloc} and provide promising results. \\
Unfortunately, in a real-world settings, 24/7 geographical coordinates sensing is not possible. Especially in indoor environments and due to battery limitations, GPS is usually turned off. Cell-ID provides geographical coordinates, and it is more frequently available, but it is too imprecise for location recognition \cite{enefpossmart}. There sometimes is even no geographical data available at all, e.g. with the Device Analyzer dataset. In these cases, we use the WiFi data for estimating the location. Based on this description, the location estimation algorithm must be able to identify location changes from a combination of sensors (sensor fusion). Furthermore, it is important to note that our focus is on the data that is being collected from users' context-sensing device (user-centric) and not a third party data such as a call detail record (CDR) \cite{hummob, idimpplpl}.\\
\vspace{-0.1cm}
 \begin{algorithm2e}[htb]
 \scriptsize 
 \KwData{$entities$,$signalType$}
 \KwResult{$results$}
 	 \If {($isWiFi$ = $signalType$)} {
			\ForAll { ($locations$ in $entities$) }  { 
					$moving$ \textleftarrow\  $contDiff$($locations$); \\
					\If {($moving$ != $\varnothing$)} {
						$results.add$($moving$) ; \
					} \ElseIf  {( $moving$ = $\varnothing$ \& \\
				           $contSim$($locations$) != $\varnothing$  )} {
							$results.add$($stationary$); \
					} \Else {
						$results.add$($unknown$); \
					}
				}  
			}	\Else { 
			 \ForAll { ($locations$ in $entities$) }  { 
				$locstate$ \textleftarrow\  $parseGPS$($locations$); \\
				\If	{($locstate$ = $\varnothing$)} {
					$locstate$ \textleftarrow\  $parseOtherSignals$($locations$); \
				}
				\If	{($locstate$ = $moving$)} {
					$results.add$($moving$) ; \
				}\ElseIf  {( $locstate = stationary$ \&
				    $contSim$($locations$) != $\varnothing$  )} { 
					$results.add$($stationary$); \
			    } \Else {
			       $results.add$($unknown$); \
			    }
			 }
		 }
 return $ results;$ 
\caption{\footnotesize Location state estimation from different signals.}  \label{alg:location}
\end{algorithm2e}
\normalsize
Algorithm \ref{alg:location} presents our location state estimate. The algorithm receives a set of \emph{entities} and \emph{signal type} as input, and it returns a list of entities with location state in \emph{results}. We call entities with location states as events. Entities are timestamped sensor/data records and signal type could be WiFi or a combination of sensors. An event includes an spatial state, start time, end time, and set of entities. \\
As the first step, the algorithm checks $signalType$, line 1. If the signal type is only WiFi, then it returns true; otherwise, there is a combination of location signals, and the algorithm continues from line 11. The $contDiff$ method at line 3, searches for a sequence of continuous WiFi BSSIDs, which have different names, if there exists a sequence, and if no WiFi BSSID has been repeated in the sequence, this is a sign of a moving event. Therefore, a \emph{moving} event is created and appended to the $results$ list, line 5. Otherwise, if there is a sequence of WiFi BSSIDs, but at least one of them is repeated (they are not unique), the $contSim$ method returns them, and a \emph{stationary} event will be created and appended to the event list ($results$), line 8. For instance, the sequence ${w_1,w_2,w_3,w_4,w_1,w_3}$ includes two repeated elements, $w_1$ and $w_3$ and the algorithm creates an \emph{stationary} event from this sequence. In line 10, if there is no WiFi signal at all, then an \emph{unknown} event will be created and appended to the $results$ list. In summary, the algorithm checks WiFi BSSID data objects that appeared sequentially and if they are not unique, then it creates \emph{stationary} events. Otherwise, if there are unique elements, the algorithm creates a \emph{moving} event, and if none of these cases exists, then it creates an \emph{unknown} event. \\
If there exists geographical coordinates, the status of the location is easily recognized. To calculate location state from geographical coordinates the algorithm checks the differences between two consecutive points and calculates the state (if it is moving or stationary). Method $parseGPS$, line 13, implements this scenario. \\
Nevertheless, in a real world settings, most of the time the GPS is turned off and there are very few GPS logs (mostly when users are navigating). In the UbiqLog dataset, most location logs will be from Cell-ID. As it has been described, it is not possible to precisely identify if the user's location has changed or just the cell tower has been switched. Therefore, the calculation should be flexible with 800 to 1000 meter accuracy \cite{discgps}. To cover this precision problem, instead of calculating the distance between two consecutive points (geographical coordinates), we calculate the distance between three consecutive points. If the distance between points 1 and 3 is more than 800 meters, then we can conclude the user is \emph{moving}. Method $parseOtherSignals$, line 15, implements location state calculation from Cell-ID data objects.\\ 
The complexity of algorithm \ref{alg:location} is linear, $O(n)$, because even if we assume all locations are Cell-IDs there is no need for a comparison between each element and its two previous ones. Therefore, in a worst case scenario, we will have a $3n$ comparison, which is still linear.\\
Afterward, we created a file for each user, which includes records of their data. Each record includes four elements: sensor name, timestamp, sensor value and location states which is a presentation of a four-tuple entity.
\subsection{Temporal Granularity}
According to \cite{time}, we do not perceive time in and of itself, but rather, we perceive changes or events in time. To be able to model human behavior, a precise machine timestamp should be transferred to a format similar to the way human perceive time. In a more technical sense, humans perceive events in relation to both location and time \cite{towsharing}. Unlike location data loss problem, all existing digital mobile and wearable devices can record the sensed information objects with a timestamp. In order to simulate human perception we have studied the literature in temporal data analysis \cite{encydb}. \\
This concept has been identified as the temporal granularity \cite{tempgran}. Temporal granularity is application-specific and therefore there is no generic solution that can be applied to all problems. Here we attempt to make a temporal granularity for the daily behavior. For instance, a user makes a telephone call to his mother in the evening, regularly. It is unlikely that he will call her every day exactly at 5:00; he could call  one day at 5:21 and another day at 4:53. As a result, we define temporal granularities based on \emph{common daily time scheduling}, and we provide an algorithm that can convert times based on the given precision. \\
\begin{algorithm2e}[h]
\scriptsize
 \KwData{$D_{in}$, $precision$}
 \KwResult{$D_{out}$ }
  //iterate through entities of a date \
  \For  { ($i$=0 ; $(i < D_{in}.e(length)$) )} { 
    // read hour and minutes of current entity \\
	 $TmpCeil \gets ceil(D_{i}(T)_{H},persicion) $ \;
	 $TmpFloor \gets floor(D_{i}(T)_{H},persicion) $ \;
	 $ T_{abs}  \gets distance(TmpCeil, TmpFloor)$ \;
	 $D_{i}(T) \gets T_{abs} $ \;
	 $D_{out}.add(D_{i}(T))$   	
   	}
  return $D_{out}$;
	\caption{\footnotesize Temporal granularity calculation.}\label{alg:tempgran}
\end{algorithm2e}
\normalsize
The algorithm we propose for temporal granularity is flexible enough to work with different timeframes, but in our experimental evaluation we define six time frames: Five minutes (for time-sensitive tasks such as attending a meeting), a quarter of an hour, half an hour, an hour, one and half hour and two hours. \\
Algorithm \ref{alg:tempgran} denotes the transformation of a timestamp. It receives a day object $D_{in}$ plus the $precision$ in minutes (i.e. 5', 15', 30', 60', 90' or 120'). $D_{in}$ contains $i$ number of entities. First, it extracts the hour and minutes from the timestamp and calculates the $ceil$ and $floor$ based on the given precision, line 3,4. The $distance$ function, line 5, simply checks the distance of the given ceil and floor and returns the smaller one. For instance, if the precision is set to 5 minutes and the given time is 11:08', the ceil would be +2 minutes, 11:10', and floor would be -3 minutes, 11:05'. The distance function returns ceil and thus the 11:08'  would be converted to 11:10'. Then this value will be substituted as a time in its related entity, line 7, and a new day object with a converted timestamp will be returned. The computational complexity of this algorithm is $O(n)$ too.\\
It is notable that this temporal similarity transformation can handle uncertainty by focusing on similar data in a perceptible time interval (i.e. a quarter of an hour, half an hour, etc.). Therefore, this model can be used to handle uncertainty originating from different time of routine behaviors.
\subsection{Motif Identification and Profiling}
\begin{figure}[htb]
\includegraphics[scale=0.8]{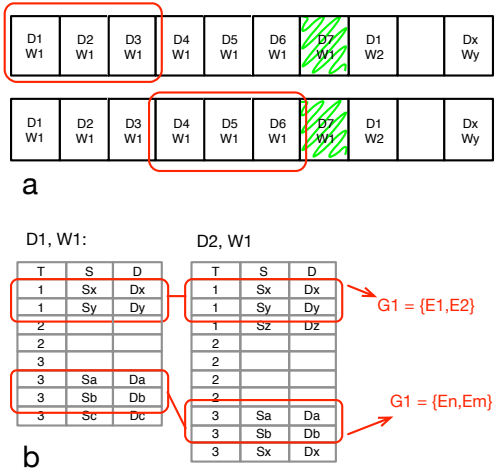} 
\vspace{-0.2cm}
\caption{\small Semantic visualization of group creation based on similarities between entities. Firgure (a) presents a sliding window with a size of three. Figure (b) presents similar entities that have been detected between two days; window size and $\theta$ both are equal to two.}\label{fig:slwin}
\vspace{-0.1cm}
\end{figure}

After the data has been transformed and its timestamp has been converted, then the similarity detection algorithm starts to build groups of similar activities. First, we introduce group creation algorithms from similar entities, and then we describe the method that builds users' profiles by extracting behavioral motifs from groups.
\begin{algorithm2e}[htb]
\scriptsize
 \KwData{$D_{ins}$, $ws$,$ \theta $}
 \KwResult{All Detected Groups in a Window}
  $grpAll , grpPrev \gets \varnothing$ \;
  $entArr , entArrNext \gets \varnothing$ \;
 \While{($(D_{ins}.hasNext)  <ws $)} { 	
	 //reading entities of current day  \\
     $entArr \gets D_{ins}.current.e $ \;
     //reading entities of next day  \;
     $entArrNext \gets D_{ins}.next.e $ \;	
	 //compare and create groups	  \;
	$entSimilar \gets compare(entArr, entArrNext, \theta)$ \;
	// collect similar entities \;
	$grpTmp.add(entSimilar)$ \;
 	\If{($grpPrevious.containsData()$)} {
 		   $grpPrevious \gets getSimilar(grpTmp,grpPrev)$\;
 		   $grpAll.add(grpPrev)$ \;
 	  } \Else {
 			$grpAll.add(grpPrev)$ \;
 	  }			
  } 
 return $groupAll$	;
 \caption{\footnotesize Group creation from similar entities.}\label{alg:group}
\end{algorithm2e}
\normalsize
Figure \ref{fig:slwin} visualizes the algorithm \ref{alg:group} we propose for group creation . The window size is set to be three; one day as a weekend will be neglected\footnote{\small{The target city of the experiment only has a one-day weekend.}}, and $\theta$ is equal to two. By comparing two days, D1,W1, with D2,W2, two groups, G1 and G2, have been extracted. For the sake of brevity we did not visualize a comparison between more than two days. Algorithm \ref{alg:group} receives input days, $D_{ins}$, window size $ws$, and minimum threshold $\theta$. In line 3, it iterates through days, $D_{ins}$, and reads entities for each day. Then it compares the current entities to the next day's entities using the $compare$ method and keeps the similar ones in a temporary group $grpTmp$, lines 9 to 11. If a previous similarity group exists, $grpPrev$, then it updates that group via the $getSimilar$ method, in line 13. This process repeats for given learning days and all similar groups in the given window size. The result will be collected and returned in the array called $grpAll$. In summary, each window returns a set of groups.\\
Collected groups are raw behavioral motifs. Since behaviors are just combinations of groups, we can add them all together to have one set that includes group objects, this set will be called $Profile$. Algorithm \ref{alg:profile} summarizes the collected groups and returns the profile object. In line 3, it iterates through objects of a given Groups array. It increases the confidence of repeat group objects and removes them from the array in line 5 to 8. Then it calculates an intersection between groups, and if the appearance of a group is more than the $\lambda$ threshold (line 10), then this group will be added to the user's profile. At the end, it returns the $Profile$ object. Both algorithms \ref{alg:group} and \ref{alg:profile} are linear, $O(n)$.\\
\begin{algorithm2e}[htb] 
\scriptsize
 \KwData{$Groups$, $\lambda$}
 \KwResult{$Profile$}
 $Profile \gets \varnothing$\;
 // finding similar groups \;
 \While{($Groups.hasNext$) } {
       // two groups are equal \;
		\If { ($Groups.next = Groups.current$)} {
			// increase the confidence of the current group
			$Groups.current.confidence + 1 $ \;
			// remove the repeated group\
			$Groups \gets remove(Groups.next)$ \;
		}
	}
  // prune groups confidence based on $\lambda$ \;
  \While {($Groups.hasNext$) } {
  	\If { ($Groups.current.confidence \geq \lambda $)} {
		$Profile.add(Groups.current)$ \;		  	 
  	 } 
  }
 return $Profile$;
\caption{\footnotesize Creating profile from behavioral motifs.}\label{alg:profile}
\end{algorithm2e}
\normalsize
As has been previously noted, groups are the unit for predicting and quantifying human behavioral dynamics. Existing works \cite{ace, mobileminer} provide association rule mining on pure contextual information. Since this work aims to identify human behavior, instead of the unique contextual information approach we propose temporal group-based contextual information.  \\
It is important to note we still cannot map these information objects onto real-life events, but our work offers a significant step toward more intuitive understanding of human behavior (especially with the temporal granularity we are using). Moreover, our approach does not rely on a unique sensor; therefore, data is extracted from multiple sensors so if a single sensor fails, its impact is insignificant. This helps mitigate the problem of uncertainty that originated from the sensor data.
\vspace{-0.2cm}
\section{Experimental Evaluation}
In order to demonstrate the utility and efficiency of our algorithms\footnote{\small{To allow full reproducibility of this paper's claims, all code and the UbiqLog dataset are available. Please contact authors for more information.}}, we consider the following four experiments:
\vspace{-0.2cm}
\begin{itemize}\itemsep1pt \parskip0pt \parsep0pt
\item{We demonstrate the scalability of the algorithm by analyzing the impact of window size (WS) on the execution time performance on two dataset, Device Analyzer and UbiqLog. Since (i) our behavioral pattern mining approach should be lightweight enough to be used in wearable and mobile devices and since (ii) these devices have limited resources, investigating the execution time is very crucial to demonstrate if our approach is scalable on these devices or not. In order to demonstrate the versatility of our algorithms, in addition to UbiqLog dataset, we test the execution time performance on 35 random users from the Device Analyzer dataset too.}
\item{We evaluate the accuracy of the motif identification algorithms based on different segments of a day and temporal granularity. In particular, we investigate the impact of changing different values of temporal granularities and day segmentation on prediction accuracy.}
\item{We examine users' behavioral changes over time. This enables us to identify the best time to run the algorithm. Due to the resource limitation of wearable and mobile devices, it is important to understand the approximate time for executing such algorithms.}
\item{Finally, we present an statistical overview on the impact of changing thresholds (i.e. $\lambda$ and $\theta$) and temporal granularities on motif identification. This will help us to identify boundaries to configure these variables.}
\end{itemize}
\vspace{-0.2cm}
\subsection{Execution Time \& Scalability} \label{exectimescal}
\begin{figure}[h]
	\hbox{\hspace{-3ex}\includegraphics[scale=0.37]{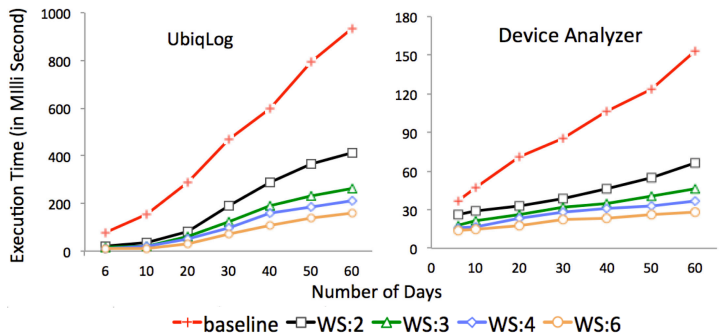}}
	 \vspace{-0.4cm}
		\caption{\small The effect of window size (WS) on the execution time performance for both UbiqLog and Device Analyzer datasets.}\label{fig:performance}
\end{figure}

Scalability is one of the novel contributions of this work and it will be done through the adoption of a sliding window, to improve the execution time performance. In particular, our algorithms must be capable of being integrated into small devices which have restricted computational resources compared to desktop computers. \\
We have analyzed the execution time performance of our algorithms with different window sizes for 60 days. The base line is not using the window and it compares each day with all other days in the dataset. Figure \ref{fig:performance} summarizes these performance changes for both UbiqLog and Device Analyzer datasets. The legend on the bottom shows the window size (WS); we have tested for window sizes 2,3,4, and 6. The result shows that increasing the window size significantly improves the performance. In other words, a smaller slope means better performance, and increasing the window size decreases the slope significantly in both datasets. Even increasing the number of days, does not affect the performance of the algorithm. 
\vspace{-0.3cm}
\\
It is notable that numbers depicted in Figure \ref{fig:performance} belong to all 35 users and an application that uses this approach will use data only for one user. Therefore, the execution time per person will be reduced significantly. However, they have been measured on a MacBook with 2.4 GHz CPU and 8 GB RAM. If the algorithm is ported to a wearable or mobile device that has limited resources, these numbers will increase based on the device's capabilities.
\vfill{}
\subsection{Temporal Granularity Impact on Accuracy \\ of Motifs Identification} \label{tempgral}
In order to evaluate the accuracy of our motifs identifications algorithms, we created a ground truth dataset by annotating a subset of randomly selected motifs. We have randomly chosen 5000 motifs and asked related users to annotate them. Users were asked to annotate if they agree with the selected motifs in their profile or not (True/False).\\
With regard to our ``Temporal Variety of Motifs \& Users Characteristics'' experiment (See section \ref{userchar}), we examine the accuracy of our algorithms using three temporal segments of a day: 0:00-8:00, 8:00-16:00 and 16:00-23:59 (we use 24 for the sake of readability). Furthermore, since our temporal granularity algorithm is flexible enough to work with different time frames, we evaluate the accuracy of motif identification using six different described temporal granularities. \\
Finally we count motifs with more than 20\% confidence as positive prediction results, and lower than 20\% confidence as negative prediction results. Table \ref{tab:acc} shows the result of analyzing the accuracy of motif prediction. The base line (first row of the Table \ref{tab:acc}) is not using temporal granularity.\\
We find that motif prediction performance is influenced by different values of temporal granularity and the segmentation of the day. The results show that identifying motifs using ``an hour'' as the temporal granularity improves the performance compared to other temporal granularities. Nevertheless, choosing 15' ,30', 90' and 120' as temporal granularity performs almost the same or slightly lower than 60' but better than 5'. This is due to the fact that five minutes is too precise for an application to model human behavior. Most of routine human behaviors that can be identified from a smartphone have one hour approximation. \\
In addition to that, we find out that the performance of the motif identification is more accurate for two segments of the day 0-8 and 8-16. This is due to the fact that 0-8 is a sleep time (very routine behavior) and 8-16 is work / school time (also a routine behavior). In contrast, 16-24 (leisure time), has a low likelihood of having routine behaviors. 
\vspace{-0.2cm}
\subsection{Temporal Variety of Motifs \& \\ Users Characteristics} \label{userchar}
\begin{figure}[h]	
	\hbox{\hspace{-3ex}\includegraphics[scale=0.5]{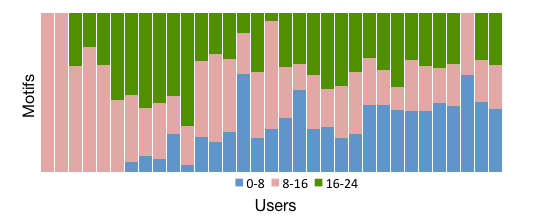}}
	\vspace{-0.3cm}		
		\caption{\small Motif distribution among users in three temporal segments.}\label{fig:motifdist}
		\vspace{-0.2cm}
\end{figure}

As has been stated previously, data mining is a resource-intensive task, and mobile and wearable devices suffer from resource weaknesses in comparison to desktop computers. This reveals the fact that data mining algorithms should not run continuously on small devices. It is important to know \emph{when to run} the these algorithms, or how often it is necessary to run our algorithms and detect new behavioral motifs. \\
To achieve this goal, we need to understand users' characteristics based on the temporal aspect of their behavioral motifs. Therefore, we analyze temporal differences among users in terms of their routine behavior. Identification of these temporal differences enables the target system to decide about the best execution time for its algorithms. For instance, users could have routine behaviors during the evening and not many routine behaviors during the day; if a system learns this, then it will execute its algorithms on evening data and not run its algorithms during the day. \\
For the implementation of this approach, again, first we segment a day into three temporal segments, which is more accurate than two divisions proposed by \cite{habitmin}. Each segment covers eight hour of a day: 0-8, 8-16 and 16-24. Figure \ref{fig:motifdist} shows motifs detected in three temporal segments for UbiqLog users. 
The stack bar plot in Figure \ref{fig:motifdist} has been ordered based on the number of motifs detected from 0:00-8:00. This figure does not visualize motifs confidence. Each motif has a confidence, and we assume motifs that have more than 20\% accuracy are highly routine motifs. 
\vspace{-0.3cm}
\begin{figure}[h]
\begin{center}
	\includegraphics[width=0.7\textwidth]{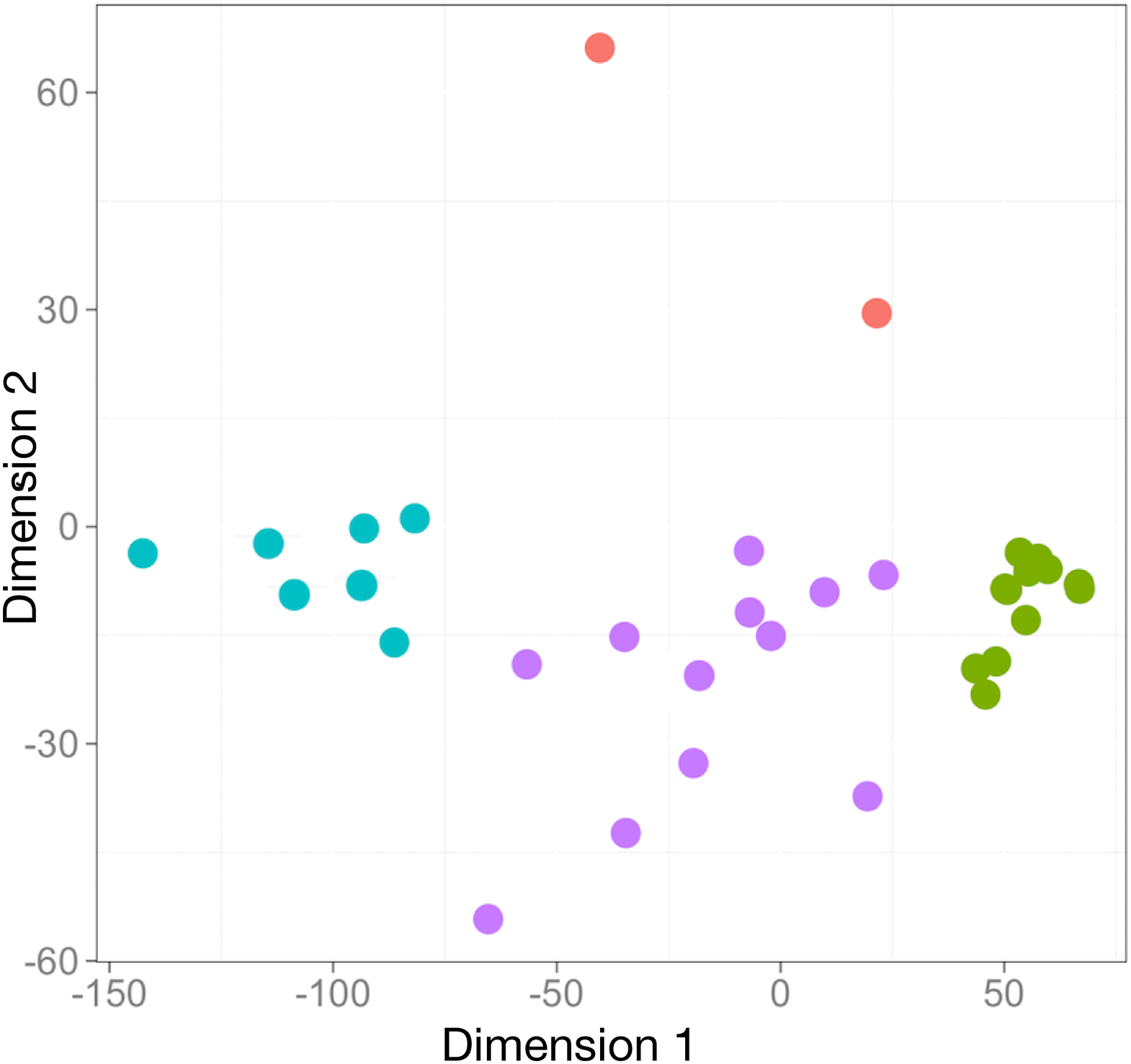}
	\vspace{-0.6cm}
	\caption{\small Multidimensional scaling of clusters of users based on the temporal distribution of their motifs. We have colored the cluster elements based on the characteristics of users. (best viewed in color)}\label{fig:mds}
\vspace{-0.3cm}
\end{center}
\end{figure}
\vspace{-0.4cm}
The second step is to identify if we can generalize users' characteristics on the described temporal segment or not. We use a topic modeling approach, latent semantic indexing \cite{lsa} (LSA), to cluster users based on their temporal motifs within their confidence. Our approach assumes users as \emph{documents} and numbers of motifs with their temporal segment plus confidence as \emph{terms}. These are terms: \texttt{0-8 \& <20\%}, \texttt{0-8 \& >20\%}, \texttt{8-16 \& <20\%}, \texttt{8-16 \& >20\%}, \texttt{16-24 \& <20\%}, \texttt{16-24 \& >20\%}.  \\
Figure \ref{fig:mds} shows a multidimensional scaling \cite{mds} we have performed on the result of our LSA clustering. The two red dots on the top are outliers, which have significantly different behaviors than other users. Our results shows three clusters. The green dots, which present higher density are users that provide fewer motifs during 16-24 with high confidence and a larger number of high confidence motifs during 8-16. The other cluster, in purple color, shows users that have an average number of motifs (with both low and high confidence) distributed among two segments: 8-16 and 16-24. Dots in blue color presents the third cluster. This cluster has more identified motifs at 0-8 temporal segment and fewer motifs in the other two segments. \\
Gaining such an insight, (i.e. the temporal segment with the highest motif discovery rate) enables a system to identify the best execution time for motif detection. In addition to that, this clustering assists the system to reduce the search space through filtering data that is not being used for motif discovery. In other words, if a system knows a user's cluster, our motif detection algorithms can be applied to only one third of the data (one temporal segment) and thus reduce the search space by one third (numerosity reduction).
\vspace{-0.2cm}
\subsection{Thresholds Effects} \label{teffect}
\begin{figure*}[ht]
\begin{floatrow}
\CenterFloatBoxes
\ffigbox{%
}{
  \hbox{\hspace{-3ex}\includegraphics[scale=0.6]{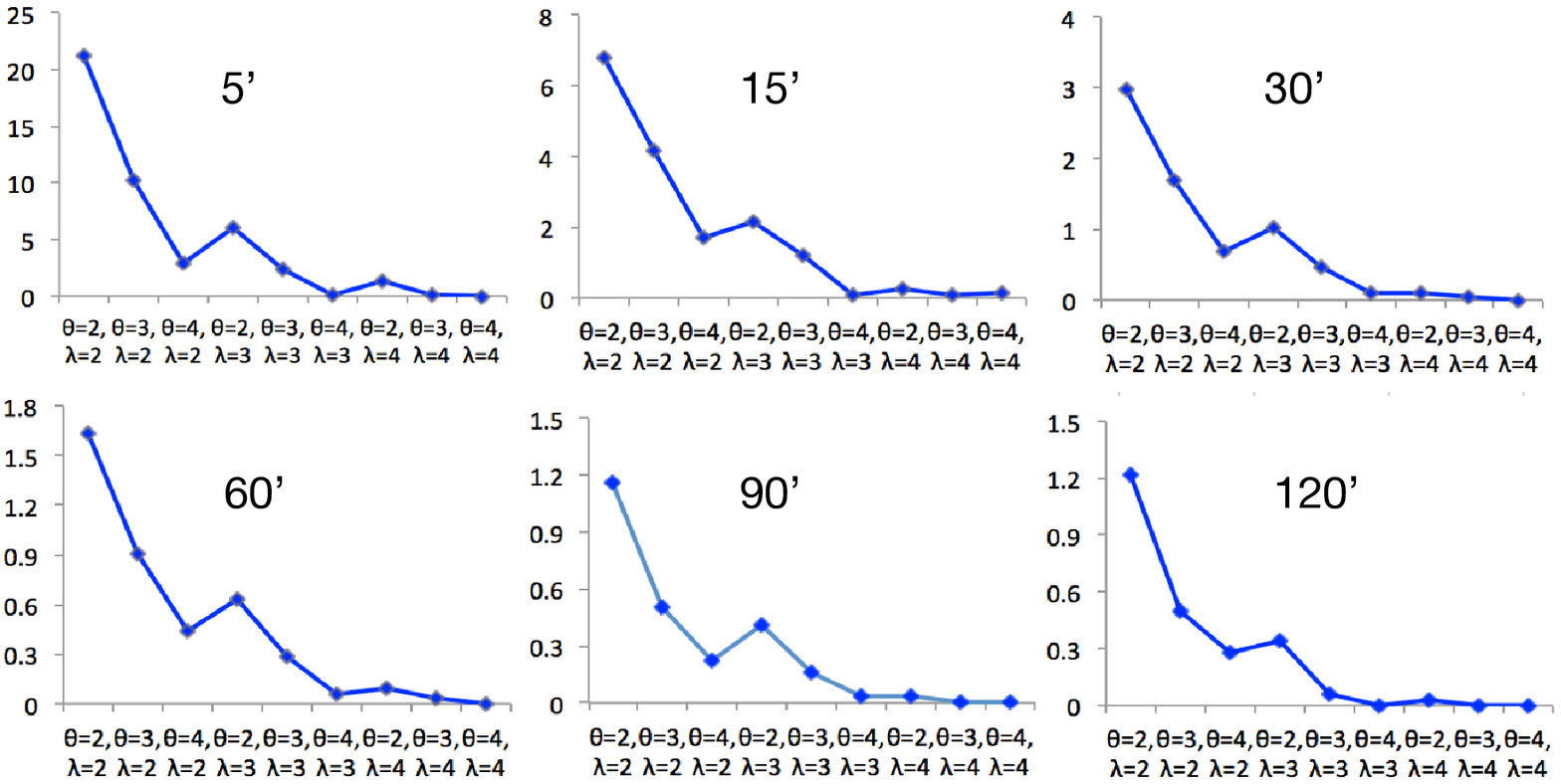}}
  \caption{\hbox{\small Average number of behavioral motif (X Axis) for each temporal granularities,} \hbox{based on different activity thresholds $\theta$ and behavioral motif confidences $\lambda$ (Y axis)}.}\label{fig:vars}
}
\capbtabbox{%
  \begin{tabular}{c|cccc} \hline
  & \multicolumn{4}{c}{Temporal Segment} \\
    & 0-8   & 8-16 & 16-24 & All  \\ \hline
0' & 0.52  & 0.56 & 0.42  & 0.48    \\ 
5'  & 0.67  & 0.60 & 0.62  & 0.62 \\
15' & 0.81  & 0.81 & 0.73  & 0.78 \\
30' & 0.88  &   0.78  & 0.74  & 0.79 \\
60' & \textbf {0.88}&\textbf {0.86} &\textbf {0.88}& \textbf {0.87} \\
90' & 0.80  &  0.77  & 0.70  & 0.75 \\
120' & 0.77  &   0.78  & 0.77  & 0.78 \\  \hline
All & 0.77  & 0.72 & 0.68  & 0.72 \\ \hline
  \end{tabular}
}{
  \caption{\small Evaluation results of motifs identification with regard to different temporal segments of a day and different temporal granularities. 0' is the baseline, i.e. not using the temporal granularity.
\label{tab:acc}}%
}
\end{floatrow}
\vspace{-0.2cm}
\end{figure*}

``Activity threshold,'' ``behavioral motif confidence'' and ``temporal granularity'' are three configurable variables. We test our motif identification and profiling algorithms with six different types of temporal granularities of 5', 15', 30', 60', 90', 120'. Figure \ref{fig:vars} shows the average number of detected behavioral motifs for each temporal precision from different activity thresholds ($\theta$) and the behavioral motif confidence ($\lambda$) in UbiqLog dataset. There is no specific suggestion for a best combinations of these variables because their usage is context and application dependent. For instance, if just estimating the location is enough for an application $\theta$=1 will be enough, but if the target application tries to identify location and the routine activities inside that location, $\theta$ should be set to two or more. \\
However, these results show that increasing both $\lambda$ and $\theta$ to more than three reduces the chance of detecting any behavioral group in the UbiqLog dataset. Nevertheless, based on the maximum number of identified motifs ($\theta$ =2) in Figure \ref{fig:vars}, we demonstrate that it is not feasible to model and predict human behavior 24 hours in a day, via smartphone. This finding is inline with \cite{gettingcloser}, which argues that the smartphone's proximity to users, restricts a 24/7 behavior observation.\\
In addition to this, Figure \ref{fig:vars} shows a possible maximum for the activity threshold. In particular, there will be very few behavioral motifs identified with a $\theta$ larger than three, and a $\lambda$ correlated with the number of identified motifs but as effective as $\theta$.
\vspace{-0.2cm}
\section{Related Work}
A major contribution of this research is a generic mobile data mining system. We claim it is generic because of its multi-sensor support and application independence. A secondary contribution is our modeling of the temporal aspect of human behavior. Moreover, we discuss algorithms for location estimation based on users' smartphone data. Therefore, we study three categories of related works: mobile data mining efforts that focus on device data collection(not 3rd party provider), temporal granularity analysis in human behavior, and location estimation from smartphone data. 
\vspace{-0.1cm}
\subsection{Mobile Data Mining}
Research that relies on collecting data from users' mobile devices is mostly application-specific and focuses on predicting one element of data (single sensor). For instance, a category of research explores activity recognition from accelerometer data \cite{unlabelsparse, activitytaxonomy, genmodautrec}. Recent approaches \cite{unlabelsparse} have tried to employ a data dictionary and use semi-supervised learning to learn human activities. This makes the data mining process light  as well as scalable for implementation on mobile devices. However, there still is no perfect solution for activity recognition even in commercial wearables, and researchers must also deal with the uncertainty problem of activity recognition techniques \cite{validity}. There are two works relevant to our research: MobileMiner \cite{mobileminer} and ACE \cite{ace}. Both studies are very similar and consider the co-occurrence patterns in human behavior via mobile phones through association rule mining. Their approach is realistic in terms of deployment, but since they use association rule mining, they are restricted to co-occurrences of more than one data object. In contrast, we identify behavior motifs and not just co-occurrences. Likewise, since we aim for human behavior detection we benefit from the temporality of behavior, and thus there is no need to have at least two data objects available for prediction (one is enough if the application uses  $\theta =1$). Another similar work is \cite{habitmin}, which extracts users' routine behavior by identifying application usage correlation with time and location. This work transforms geographical coordinates based on the time of the day to ``work'' or ``home''. Our location transformation is more accurate than this transformation, and we include precise time of the day while transforming the location.
\vspace{-0.1cm}
\subsection{Human Behavior Temporal Granularity }
As has been stated previously, our work tries to digitally map timestamps for human activities onto human temporal perception. The term temporal granularity has been introduced by \cite{tempgran},  and is different from temporal abstraction. It is notable that temporal abstraction is the process of converting high-dimensional timestamp data to low-level qualitative descriptions of time \cite{tempabssurv} and  has been introduced by \cite{kbabs}. Temporal granularity specifies the temporal qualification of a set of data, similar to its use in the temporal qualification of statements in natural languages. \\
We review temporal granularity models that are being used for human behavior analysis. There is a limited number of works that consider how to apply temporal granularity to human behavioral data. One of the earlier works, \cite{layrephactrec}, proposes a method to  analyze human activities in the office via a probabilistic representation for inferring temporal granularity. Our goal is not to infer temporal granularity, but we benefit from this concept to mine patterns of human behavior. The work most similar to ours is \cite{minindlochist},  which focuses on mining users' daily location patterns via trajectory mining and  defines the temporal granularity as a day. As has been stated previously, \cite{habitmin} is another approach for identifying daily behavior and tries to match the daily location of users to application they use. They converted a day into two segments (8-18 and 18-8) and model application usage in each segment.
\vfill{}
\subsection{Location Estimation from Smartphone}
There are several research benefit from smartphone location logs, i.e. GPS, WiFi, Cell-ID, to identify locations of interest and daily movement patterns. Reality mining \cite{eigenbehav}, is one of the first effort toward identifying behavior from smartphone contextual data. They have created a benchmark dataset that is still using widely by research in this area \cite{probappmin, unsupframe, autrec}. These researches use Reality Mining location data and mine patterns of daily location changes. Recently the uncertainty of a realistic deployments have been taken into account and there some works tries to support uncertainty while mining for location data originated from smartphone unreliable sensors too \cite{autrec}.  Semantically, location is the most valuable information in digital human behavior identification, and therefore these studies map location onto human behavior. We believe human behavior is not just based on changes in location, and studies should include activities that are happening in the location too. Therefore, our interpretation of human behavior is different from those interpretations. Since our research can use all existing sensors on the device, it can be extended to any type of human behavior analysis application. In other words, we benefit from a \emph{combination of sparse information sources} and not just one information source.
\vspace{-0.2cm}
\section{Conclusion \& Future Work}
In this paper, we have proposed a \emph{scalable} approach for daily behavioral pattern mining from multiple information sources. This work benefits from two \emph{real-world} datasets and users who use different smartphone brands. We use a novel \emph{temporal granularity} transformation algorithm that makes changes on timestamps to mirror the human perception of time. Our \emph{behavioral motif} detection approach is generic and not dependent on a single source of information; therefore, we reduce the risk of uncertainty by relying on a combination of sensors to identify behavioral motifs and patterns. Results of experimental evaluation show that using sliding window significantly decreases the execution time. Moreover, converting raw timestamps to temporal granularities increases the accuracy of motifs identification, which is influenced by different values of temporal granularity and the segment of a day. In particular, we find out that setting temporal granularity to one hour has the highest motifs identification accuracy. Finally, quantifying users' characteristic based on their temporal motif distributions results in three group of users. This finding assists the system to identify appropriate run time.\\
In future work, we are going to extend the concept of granularity by making it dynamic and not a using similar granularity for both start time and end time. Additionally, we plan to substitute our sliding window approach with damped window, which is a weighted based comparison. Average motif life time, will be extracted, and the diminishing weight of damped window will be calculated based on the average motifs life time. \\
\textbf{Acknowledgments:} We acknowledge Michael Pazzani, Eamonn Keogh and Vassilis Tsotras for their helpful feedback and comments on this research. 
\vspace{-0.2cm}
\section*{}
\vspace{-0.4cm}
\bibliographystyle{abbrv}
\small{\bibliography{refs}}
\end{document}